\documentclass[a4paper,11pt]{article}
\usepackage{pos}
\usepackage{slashed}

\newcommand{\mev}{\text{ MeV}}

\newcommand{\mdm}{M_\text{DM}}
\newcommand{\hc}{\text{h.c.}}
\newcommand{\chidm}{\chi_\text{DM}}
\newcommand{\vrel}{v_\text{rel}}

\newcommand{\Leff}{\Lambda_{\text{UV}}}

\newcommand{\omix}{\mathcal{O}_{0}}

\newcommand{\ochgd}{\mathcal{O}_{+}}

\title{Electroweak Multiplets as Dark Matter candidates: \\ A brief review}
\author[a,b]{Paolo Panci}

\affiliation[a]{Dipartimento di Fisica E. Fermi, Università di Pisa,\\
  Largo Bruno Pontecorvo 3, I-56127 Pisa, Italy}

\affiliation[b]{INFN, Sezione di Pisa,\\
Largo Bruno Pontecorvo 3, I-56127 Pisa, Italy}

\emailAdd{paolo.panci@unipi.it}

\abstract{

I provide a thorough review of the theoretical and experimental status of ElectroWeak multiplets as Dark Matter candidates, serving as the prototype of Weakly Interacting Massive Particles (WIMPs) Dark Matter. Specifically, the examination includes both real SU(2) representations with zero hypercharge and complex ones with $Y\neq 0$. For the first time, all calculable thermal masses for scalar and fermionic WIMPs are computed, incorporating significant non-perturbative non-relativistic effects such as Sommerfeld enhancement and the formation of WIMP bound states. WIMP masses of few hundred TeV are shown to be compatible both with $s$-wave unitarity of the annihilation cross-section, and perturbativity.  Additionally, a strategy is outlined for probing these scenarios in the next generation of experiments.

}

\FullConference{Corfu Summer Institute 2023 "School and Workshops on Elementary Particle Physics and Gravity" (CORFU2023)\\
 23 April - 6 May , and 27 August - 1 October, 2023\\
Corfu, Greece\\}

\begin{document}
\maketitle

\section{Introduction}

Weakly Interacting Massive Particles (WIMPs), typically characterized by masses around the weak scale (approximately 100 GeV to 1 TeV) and interacting via the weak force (or a weak-like force), are motivated by two distinct theoretical rationales. Firstly, they are believed to have been produced in the appropriate abundance via the thermal freeze-out mechanism during the Early Universe, a phenomenon sometimes referred to as the ``WIMP miracle''. Secondly, there is an expectation that new physics would manifest at the weak scale, potentially resolving the hierarchy problem of the Standard Model, with WIMP Dark Matter particles emerging as a byproduct. It is crucial to emphasize that these two motivations are independent of each other. Even if new physics (e.g.~in the form of supersymmetry) does not materialize at the weak scale, the thermal production mechanism for WIMPs remains compelling. Therefore, while WIMPs may not be as prominent as they were decades ago, they continue to represent a well-motivated Dark Matter (DM) candidate.

\medskip

A particularly intriguing possibility within this framework, due to its minimality and predictive power, is that DM consists of the lightest neutral component of an ElectroWeak (EW) multiplet with size $n$ and hypercharge $Y$. By imposing general requirements applicable to all DM candidates, one can determine the quantum number of a given multiplet such that the model's only free parameter is the mass of the neutral component. Matching this mass to the measured value of the DM abundance today, $\Omega_{\rm DM} h^2 = 0.11933 \pm 0.00091$~\cite{Planck:2018vyg}, enables the unique determination of the mass of the $n$-plet. These mass predictions serve as crucial inputs for evaluating the ability of future experimental programs to fully test the EW nature of WIMP DM. 

\medskip
This review is primarily based on Refs.~\cite{Bottaro:2021snn, Bottaro:2022one}, and it follows this structure: Sec.~\ref{sec:EWmultiplets} provides a concise overview of the EW WIMP paradigm. In Sec~\ref{sec:WIMPmass}, we delve into the key aspects of our freeze-out computation, including discussions on the unitarity bound and assessments of the theory uncertainties related to computing the thermal masses. These initial sections lay the groundwork for a comprehensive understanding of the primary signatures in direct detection (Sec.~\ref{sec:DD}), indirect detections (Sec.~\ref{sec:ID}), and collider production (Sec.~\ref{sec:Colliders}).

\section{Which Electroweak Multiplets}
\label{sec:EWmultiplets}

In general, EW multiplets can be introduced as prototypes of WIMPs through two main approaches: $i)$ Top-down approach: In this approach, EW multiplets naturally emerge in various Beyond the Standard Model (BSM) theories that primarily aim to address the naturalness problem of the EW scale. Examples include models of Supersymmetry with heavy scalars~\cite{Wells:2003tf}, from the first split-SUSY proposals~\cite{Arkani-Hamed:2004ymt, Giudice:2004tc} up to more recent constructions inspired by the anthropic principle (see~\cite{DEramo:2014urw} and references therein); $ii)$ Bottom-up approach: Here, EW multiplets are selected by imposing general requirements that all DM candidates must fulfill, without specifying the underlying UV Theory. In this review, we adopt a bottom-up approach.

\medskip
On a more specific level, following the same spirit of the original minimal DM papers~\cite{Cirelli:2005uq,Cirelli:2007xd,Cirelli:2009uv,Hambye:2009pw,DelNobile:2015bqo}, we expand the matter field of the SM by introducing a single EW multiplet with multiplicity $n$ and hypercharge $Y$. For both real and complex representations, we determine the quantum number of the multiplet by imposing four general requirements outlined as follows:

\begin{itemize}

\item[$\diamond$] Neutrality: DM must be the neutral component of the multiplet. This requirement is crucial because DM cannot possess an electric charge of order one. 

\item[$\diamond$] Stability: The electrically neutral component must be stable on cosmological scale. Hence, it must be the lightest state of the multiplet that can then be stabilized either by a discrete symmetry (e.g., matter-parity in supersymmetry) or by accidental one (multiplets with $n \geq 5$ are accidentally stable~\cite{Cirelli:2005uq, Bottaro:2021snn, Bottaro:2022one}).
\item[$\diamond$]  Not excluded by direct detection: The electrically neutral component must not be coupled at tree-level with the $Z$-boson.
\item[$\diamond$]  Perturbativity: This criterion serves to establish the maximum allowed value of $n$ for a given hypercharge $Y$ of the multiplet, thereby ensuring  the model's calculability and predictiveness. 

\end{itemize}

Requiring the neutral DM component to be embedded in a representation of the EW group imposes that the charge of a given state of the multiplet is $Q=T_3+Y$, where $T_3=\text{diag}\left(\frac{n+1}{2}-i\right)$ with $i=1,\dots, n$. At this level, we can distinguish two classes of WIMPs: $i)$ real EW representations with $Y=0$ and odd $n$; $ii)$ complex EW representations with arbitrary $n$ and $Y=\pm\left( \frac{n+1}{2}-i\right)$ for $i=1,\dots, n$. In the following we descrive the main properties of these two classes of EW DM.

\subsection{Real WIMPs}\label{sec:Realmultiplets}

At the renormalizable level, the minimal Lagrangians for real WIMPs are:
\begin{align}
\mathscr{L}_\text{s} &= \frac{1}{2}\left(D_\mu \chi\right)^2 - \frac{1}{2}M_{\chi}^2\chi^2 - \frac{\lambda_H}{2}\chi^2 \vert H\vert^2 - \frac{\lambda_\chi}{4}\chi^4\label{eq:scalarWIMP} \ , \\
\mathscr{L}_\text{f} & = \frac{1}{2}\chi \left (i\bar{\sigma}^{\mu} D_\mu - M_{\chi}\right)\chi \label{eq:fermionWIMP} \ ,
\end{align}
for scalars and fermions, respectively, where $D_\mu=\partial_\mu-i g_2 W_\mu^a T^a_\chi$ represents the covariant derivative, and $T^a_\chi$ are generators in the $n$-th representation of SU(2). The Lagrangian for the real scalar in Eq.~\eqref{eq:scalarWIMP} also includes quartic self-coupling and Higgs-portal interactions at the renormalizable level. The latter is constrained from above by direct detection limits and contributes negligibly to the annihilation cross-section.

Real WIMPs are particularly interesting because they straightforwardly satisfy, at the renormalizable level, the second and third requirements introduced in Sec.~\ref{sec:EWmultiplets}: $i)$ By construction, they do not couple at tree-level with the $Z$-boson. Therefore, they avoid the very stringent bounds from Dark Matter (DM) direct detection; $ii)$ The neutral component is automatically the lightest state of the multiplet. Indeed, the neutral component and the component with charge $Q$ of the EW multiplet are split by radiative contributions from EW gauge boson loops. In the limit $m_W \ll M_{\text{DM}}$, these contributions are non-zero and independent of $M_{\chi}$. This fact can be understood by computing the Coulomb energy of a charged state at a distance $r\gtrsim 1/m_W$ or the infrared mismatch (regulated by $m_W$) between the self-energies of the charged and neutral states. The latter can be easily computed at 1-loop~\cite{Cheng:1998hc,Feng:1999fu,Gherghetta:1999sw},
\begin{equation}
M_Q - M_0  \equiv \Delta M_Q^{\text{EW}}=  \delta_g Q^2 \simeq \frac{Q^2\alpha_{\text{em}}m_W}{2(1+\cos\theta_W)} \ ,\label{eq:splitting}
\end{equation}
where  $\delta_g=(167\pm 4)\mev$. Here,  the uncertainty dominated by 2-loop contributions proportional to $\alpha_2^2m_t/16\pi$. As one can see, this mass difference is always positive, ensuring that the neutral component is always the lightest.

\medskip
Finally from the last requirement about the perturbative unitarity of the annihilation cross section one can select the maximal size of the multiplets. We  find that the largest calculable SU(2) $n$-plet at LO is the 13-plet, which is as heavy as 350 TeV. Stronger requirements about the perturbativity of the EW sector up at high scales can further lower the number of viable candidates. Further details can be found in Ref.~\cite{Bottaro:2021snn}. 

\smallskip
Famous candidates belonging to this class of EW multiplets include the supersymmetric Wino, which is a Majorana triplet under SU(2), and the Minimal DM 5-plet.
	
\subsection{Complex WIMPs}\label{sec:Complexmultiplets}
For complex representations, the phenomenology changes substantially only when we consider multiplets with hypercharge different from zero, albeit at the cost of losing minimality. Here, we specifically concentrate on the fermionic case, reserving the discussion about complex scalar WIMPs for the Appendix of Ref.~\cite{Bottaro:2022one}. The minimal Lagrangian for a fermionic complex WIMP with $Y\neq 0$ is as follows:
\begin{align}
&\mathscr{L}_{\text{D}} =\overline{\chi} \left (i\slashed{ D}-M_\chi\right)\chi +\frac{y_0}{\Leff^{4Y-1}}\mathcal{O}_0 + \frac{y_+}{\Leff}\mathcal{O}_{+}+\hc\ ,\notag\\
&\omix=\frac{1}{2(4Y)!}\left(\overline{\chi}(T_\chi^a)^{2Y}\chi^c\right)\left[(H^{c\dagger})\frac{\sigma^a}{2} H\right]^{2Y}\ ,\label{eq:ren_lag}\\ 
&\ochgd=-\overline{\chi}T_\chi^a\chi H^\dagger\frac{\sigma^a}{2} H\ ,\notag
\end{align}

The main difference with respect to real WIMPs is that the renormalizable Lagrangian alone is insufficient to satisfy all the requirements introduced in Sec.~\ref{sec:EWmultiplets} without incorporating higher-order dimensional operators. In Eq.~\eqref{eq:ren_lag}, we only present  the minimal set of UV operators necessary to render the DM model viable. We will now illustrate the physical consequences of $\mathcal{O}_0$ and $\mathcal{O}_+$.   

\smallskip
The non-renormalizable operator $\omix$ is indispensable for removing the sizable coupling of the neutral component $\chi_N$ of the EW multiplet to the $Z$ boson.
\begin{equation}
\label{eq:Z_vectorial}
\begin{aligned}
    &\mathscr{L}_Z=\frac{ieY}{\sin\theta_W\cos\theta_W}\overline{\chi}_N\slashed{Z}\chi_N\ .
\end{aligned}
\end{equation}
This coupling would result in an elastic cross section with nuclei that is already excluded by many orders of magnitude by present Direct Detection experiments~\cite{Goodman:1984dc}. After EW symmetry breaking,  $\omix$ induces a mixing between $\chi_N$ and $\chi_N^c$. Upon replacing the Higgs with its Vacuum Expectation Value, $(H^{c\dagger})\frac{\sigma^a}{2} H$ is non-zero only if we select $\sigma^a=\sigma^+$, thus yielding the new (pseudo Dirac) mass term in the Lagrangian:
\begin{equation}
\begin{aligned}
\label{eq:neutral_splitting}
   &\mathscr{L}_m=M_\chi\overline{\chi}_N\chi_N+\frac{\delta m_0}{4}\left[\overline{\chi}_N\chi_N^c +\overline{\chi^c}_N\chi_N\right],\\
    &\delta m_0 = 4y_0 c_{nY0}\Leff\left(\frac{v}{\sqrt{2}\Leff}\right)^{4Y}\ .
    \end{aligned}
\end{equation}
$c_{nYQ}=\frac{1}{2^{Y+1}(4Y)!}\prod_{j=-Y-|Q|}^{Y-1-|Q|}\sqrt{\frac{1}{2}\left(\frac{n+1}{2}+j\right)\left(\frac{n-1}{2}-j\right)}$ contains the normalization of $\omix$ and the matrix elements of the generators. The mass eigenstates are Majorana fermions, $\chi_0$ and $\chidm$, with masses $M_0=M_\chi+\delta m_0/2$ and $M_{\text{DM}}=M_\chi-\delta m_0/2$, respectively, whose coupling to the $Z$ boson is given by:
\begin{equation}
\label{eq:Z_nonvectorial}
\begin{aligned}
    %&\mathcal{L}_S^Z=\frac{ieY}{\sin\theta_W\cos\theta_W}\chi_0^\dagger\partial_\mu\chidm Z^\mu\\
    &\mathscr{L}_Z=\frac{ieY}{\sin\theta_W\cos\theta_W}\overline{\chi}_0\slashed{Z}\chidm\ .
\end{aligned}
\end{equation}
In this case, the $Z$-mediated scattering of DM onto nucleons is no longer elastic. The process becomes kinematically forbidden if the kinetic energy of the DM-nucleus system in the center-of-mass frame is smaller than the mass splitting:

\begin{equation}
    \frac{1}{2}\mu \vrel^2<\delta m_0\ ,\quad \mu=\frac{\mdm m_N}{\mdm+m_N}\ ,
\end{equation}
where $m_N$ is the mass of the nucleus, $\mu$ is the reduced mass and $\vrel$ is DM-nucleus relative velocity. 

\smallskip
The non-renormalizable operator $\ochgd$ in Eq.~\eqref{eq:ren_lag}  is, on the other hand, indispensable to ensure that the DM is the lightest state in the EW multiplet for all $n$-plets where the hypercharge is not maximal. Indeed, EW interactions induce at 1-loop mass splittings between the charged and neutral components of the EW multiplet, which in the limit $m_W\ll M\chi$ are given by~\cite{Cheng:1998hc,Feng:1999fu,Gherghetta:1999sw}:
\begin{equation}
\label{eq:charged_split_gauge}
    \Delta M_Q^{\text{EW}}= \delta_g \left(Q^2+\frac{2YQ}{\cos\theta_W}\right)\ ,
\end{equation}
This implies that negatively charged states with $Q=-Y$ are compelled to be lighter than the neutral ones due to EW interactions. Notable exceptions include odd-$n$ multiplets with $Y=0$ (see Sec.~\ref{sec:Realmultiplets}), and all multiplets with maximal hypercharge $|Y_{\max}|=(n-1)/2$, where negatively charged states are absent. For these multiplets, setting $y_+=0$ would be the minimal and phenomenologically viable choice. Further details can be found in Ref.~\cite{Bottaro:2022one}.

\medskip
Ultimately, the requirement of perturbativity leads us to constrain the maximum value of $n$ for a given hypercharge. Specifically, for $Y=1/2$, perturbative unitarity sets the maximum $n$ to be equal to 12, which is as heavy as 253 TeV. For $Y=1$, the perturbativity of the mass splitting only allows for $n=3$ and $5$. For $Y>1$, the theory becomes unstable.

\smallskip
In summary, at the cost of losing minimality, we can formulate other benchmark WIMP models that are not excluded, with a collider phenomenology substantially different from that expected for real representations. Famous candidates belonging to this class of EW multiplets include, for instance, the supersymmetric Higgsino, which is a pseudo Dirac doublet under SU(2) with $Y=1/2$.
	
\section{WIMP Cosmology}\label{sec:WIMPmass}
After selecting the quantum number of the WIMPs in Sec.~\ref{sec:EWmultiplets}, the only free parameter of the model for a given $n$ and $Y$ is the mass of $\chi_0$, which we compute by ensuring that the multiplets are produced via thermal freeze-out. For $2\to 2$ processes, the relevant factor determining the DM comoving abundance $Y_{\mathrm{DM}}$ is the annihilation cross-section. The evolution of $Y_{\mathrm{DM}}$ as a function of $z=\mdm/T$ is described by the Riccati equation:

\begin{equation}
\label{eq:single_boltzmann}
\frac{\mathrm{d}Y_{\mathrm{DM}}}{\mathrm{d}z}=-\frac{\langle\sigma_{\mathrm{eff}}\vrel\rangle s}{Hz}(Y_{\mathrm{DM}}^2-Y_{\mathrm{DM, eq}}^2)\ ,
\end{equation}
where represents Hubble's rate, $s$ denotes the entropy density, and $Y_{\mathrm{DM,eq}}$ stands for the equilibrium comoving abundance. Here, $\langle\sigma_{\mathrm{eff}}\vrel\rangle$ denotes the thermally averaged annihilation cross-section. For EW multiplets, the primary annihilation channels consist of all EW gauge bosons. In the SU(2) invariant limit, the thermally averaged hard cross-section is expressed as:
\begin{equation}
 \langle \sigma_{\mathrm{eff}} \vrel\rangle_0 = \left\{\begin{array}{l}
\displaystyle \frac{\pi \alpha_2^2 \left(2n^4-8n^2+6\right)+32Y^4  \alpha_Y^2 +16Y^2 \alpha_2\alpha_Y \left(n^2-1\right)}{8 g_\chi \mdm^2} \ , \quad \mbox{scalars} \\
\displaystyle \frac{\pi \alpha_2^2 \left(2n^4+17n^2-19\right)+4Y^2 \alpha_Y^2 \left(41+8Y^2\right)+16Y^2 \alpha_2 \alpha_Y \left(n^2-1\right)}{16 g_\chi \mdm^2} \ ,  \quad \mbox{fermions}  \\
  \end{array}\right.
\end{equation}
where   $g_\chi=4n$ for complex representations and $g_\chi=2 n$ for real ones. This estimate is correct because during freeze-out, the temperature is higher than the mass splitting $\Delta M_Q^{\text{EW}}$, so assuming that SU(2) is unbroken is a reasonable approximation. However, it is inaccurate. Indeed, since $\mdm \gg m_W$, we expect the weak interaction to behave as a long-range force, leading to significant non-perturbative, non-relativistic effects such as the Sommerfeld enhancement and bound state formation, which could substantially amplify the tree-level estimate. Taking these effects into account, the hard cross section $ \langle \sigma_{\mathrm{eff}} \vrel\rangle_0$ has to be replaced with

\begin{equation}\label{sigmaeff}
\langle\sigma_{\mathrm{eff}}\vrel\rangle\equiv S_{\text{ann}}(z)+\sum_{B_J}S_{B_J}(z) \ .
\end{equation}
%\end{widetext}
where we have defined the effective cross-section as the sum of the direct annihilation processes, $S_{\text{ann}}$, and the ones which go through Bound State Formation (BSF), $S_{B_J}$. 

\subsection{Sommerfeld Corrections}
On a more specific level, the Sommerfeld correction can be computed by solving the Schrödinger equation in presence of a long-range EW potential which deform the wave functions of the incoming particles. In Eq.~\eqref{sigmaeff} the direct annihilation process can be factorized as
\begin{equation}
S_{\text{ann}}=\sum_I \langle S_E^I\sigma_{\text{ann}}^I\vrel\rangle\ ,
\end{equation}
where $\sigma_{\text{ann}}^I$ is the hard cross-section for a given isospin channel $I$ and $S_E^I$ is the Sommerfeld enhancement (SE) of the Born cross-section. In the limit of small relative velocity between the DM particles (but larger than $m_W/M_\chi$), the SE factor can be approximated as
\begin{equation}
S_E^I \approx \frac{2\pi \alpha_{\mathrm{eff}}}{\vrel}\ ,\quad \text{where}\quad \alpha_{\mathrm{eff}} \equiv  \frac{I^2+1-2n^2}{8} \alpha_2 \ .\label{eq:sommi}
\end{equation}
The SE correction is important for all multiplets, it becomes more pronounced at lower relative velocities and saturates when $v_{\text{rel}}\lesssim m_W/M_\chi$ due to finite mass effects\footnote{The modification of the behavior of the SE due to finite mass effects are included in our full computation (see Ref.~\cite{Cassel:2009wt} for explicit formulas).}. While this saturation is not crucial for calculating the thermal masses, it is of utmost importance for the phenomenology of DM indirect detection.

\subsection{Bound States formation}
Besides the Sommerfeld correction, if the incoming particle system possesses sufficiently high kinetic energy, particle-antiparticle pairs can combine to form a bound state, emitting an EW gauge boson. The energy of this emitted gauge boson is approximately equal to the binding energy of the BS, expressed as $E_{B_I} \simeq {\alpha_\text{eff}^2 M_\chi}/{4n_B^2}-\alpha_{\text{eff}} m_W$, where $n_B$ denotes the BS energy level, $\alpha_{\rm eff}$ represents the effective weak coupling as defined in Eq.~\eqref{eq:sommi}, and corrections of order $m_W^2/M_\chi^2$ are disregarded. In the non-relativistic limit and at the leading order in gauge boson emission, the  BSF process 
\begin{equation}
\chi_i + \chi_j \rightarrow \text{BS}_{i'j'} +V^a 
\end{equation}
is encoded in the effective dipole Hamiltonian described in Ref.~\cite{Mitridate:2017izz,Harz:2018csl}, which governs the dynamics of the BSs. Once BSs are formed, they promptly decay into SM particles, further augmenting the hard cross section $\langle \sigma_{\mathrm{eff}} \vrel\rangle_0$. Similar to the SE, we can factorize the BSF processes as:
\begin{equation}
S_{B_J}=\sum_{I,l}\langle S_E^I S_{B_J}^{I,l}\rangle R_{B_J}\ ,\label{eq:rbs}
\end{equation}
where $S_{B_J}^{I,l}$ denotes the ``hard" BSF cross-section of the state $B_J$, originating from a free state with isospin $I$ and angular momentum $l$, multiplied by the SE factor of that specific isospin channel as defined in Eq.~\eqref{eq:sommi}. Detailed expressions for this can be found in Ref.~\cite{Mitridate:2017izz,Harz:2018csl}. Conversely, $R_{B_J}$ represents the effective annihilation branching ratio into SM states, contingent upon the intricate dynamics of the BS (such as ionization rate $\langle\Gamma_{B_I\text{,break}}\rangle$, annihilation rate into SM states $\langle\Gamma_{B_I\text{,ann}}\rangle$, and decay width into other bound states $\langle\Gamma_{B_I\rightarrow B_J}\rangle$). Notably, $R_{B_J}$ converges toward 1 as the temperature of the plasma descends below the binding energies of the BSs involved in the decay chains.

In the limit where either the BS decay or the annihilation rate significantly surpasses $H$, and considering a single bound state, $S_{B_J}$ takes a rather straightforward form
\begin{equation}
\label{eq:singleBS}
S_{B_J}=\langle\sigma_{B_I} \vrel\rangle \left(1+\frac{g_\chi^2 \mdm^3 \langle\sigma_{B_I} \vrel\rangle}{2g_{B_{I}}\Gamma_{\text{ann}}}\left(\frac{1}{4\pi z}\right)^\frac{3}{2}e^{-z \frac{E_{B_I}}{\mdm}}\right)^{-1} \ , 
\end{equation}
where $\langle\sigma_{B_I} \vrel\rangle$ represents the thermal average of the cross-section for bound-state formation, and $g_{B_{I}}$ accounts for the number of degrees of freedom of the bound state $B_I$. Eq.~\eqref{eq:singleBS} applies to $1s_I$ and $2s_I$ bound states with $I\leq 5$. The latter, once formed, directly annihilate into pairs of SM vectors and fermions, with rates $\Gamma_{\text{ann}}\simeq \alpha_{\text{eff}}^5/n_B^2 M_\chi$. These bound states collectively contribute to over 50\% of the BSF cross-section. More intricate examples of bound state dynamics are elaborated in~\cite{} (where the case of the EW $7$-plet is explicitly shown). While the effect of BSF has already been computed for the fermionic $5$-plet in Ref.~\cite{Mitridate:2017izz}, here, we include it for the first time for all the real WIMP candidates with $n\geq 7$.

\begin{figure}
\centering
\includegraphics[width=0.65\textwidth]{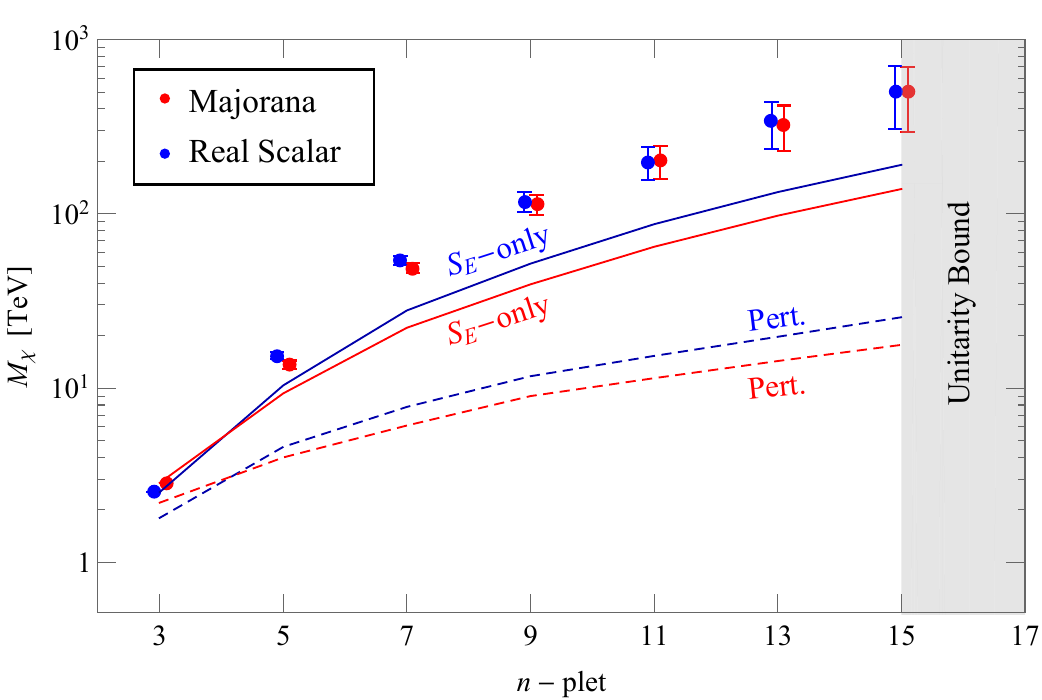}
\caption{This plot summarizes the thermal masses against the size of the multiplets for real representations, incorporating both SE and BSF. Majorana fermions are indicated in red, while results for real scalar WIMPs are in blue. Solid lines represent thermal masses with SE, while dashed lines depict masses for the hard annihilation cross-section. The gray shaded region denotes exclusion by $s$-wave perturbative unitarity, incorporating BSF}
\label{fig:summary}
\end{figure}

\subsection{Results}
We solve the Boltzmann equation, considering both SE and BSF, to provide the most accurate estimation of the thermal mass for a given $n$ and $Y$. This section summarizes our findings. 

\smallskip
Fig.~\ref{fig:summary} displays the main results for the real representations in the plane of thermal mass as a function of $n$. The blue and red colors correspond to the real scalar and Majorana fermion cases, respectively. Dashed lines represent the results considering only the hard cross section $\langle \sigma_{\mathrm{eff}} \vrel\rangle_0$. The solid lines are obtained with Sommerfeld enhancement alone, while the dots represent the full results, including the formation of BSs. As observed, neglecting non-perturbative effects leads to an inaccurate estimation of the thermal mass in all cases. This discrepancy becomes more significant for larger EW multiplets ($n\geq 5$), where the relative effect of BS  dynamics on the total cross-section increases. Approaching the unitarity bound, the error on the WIMP mass grows proportionally to the enhancement of the next-to-leading order contributions, as estimated in Ref.~\cite{Bottaro:2021snn}. Similar results are obtained for complex representations, with further details provided in~\cite{Bottaro:2022one}.

\medskip
In summary,  for $n$ bigger than 5, WIMP masses consistently surpass tens of TeV, rendering them beyond the reach of any prospective colliders. Thus, the primary means of probing these candidates is through direct and indirect dark matter detections. Conversely, for $n$ equal to or less than 5, the prospect of a potential futuristic muon collider emerges as a viable option, offering robust insights into the EW nature of dark matter.

\section{WIMP phenomenology}\label{sec:WIMPpheno}
With the computation of thermal masses completed, the phenomenology of EW multiplets becomes entirely predictable within the framework of EW theory. Consequently, we can now employ standard strategies to test the EW nature of DM. This will serve as the focus of the final part of this review.   

\subsection{Direct detection}\label{sec:DD}
Concerning direct detection, it is worth noting that all the EW multiplets, by construction, lack tree-level coupling with the $Z$ boson. Therefore, their interaction with light quarks and gluons occurs solely through one-loop and two-loop diagrams, respectively. 

\begin{figure}
\centering
\includegraphics[width=0.49\textwidth]{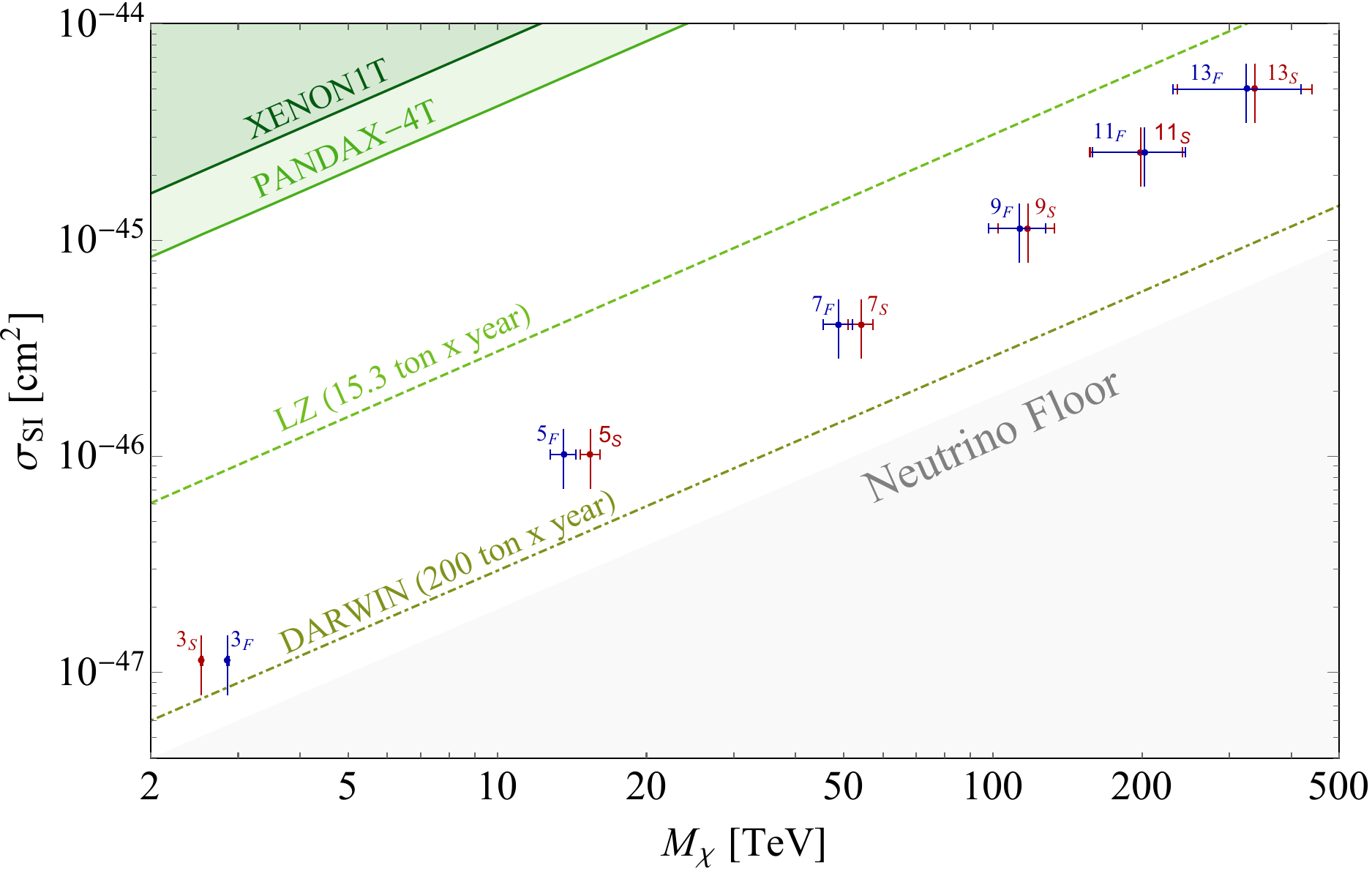} \ 
\includegraphics[width=0.49\textwidth]{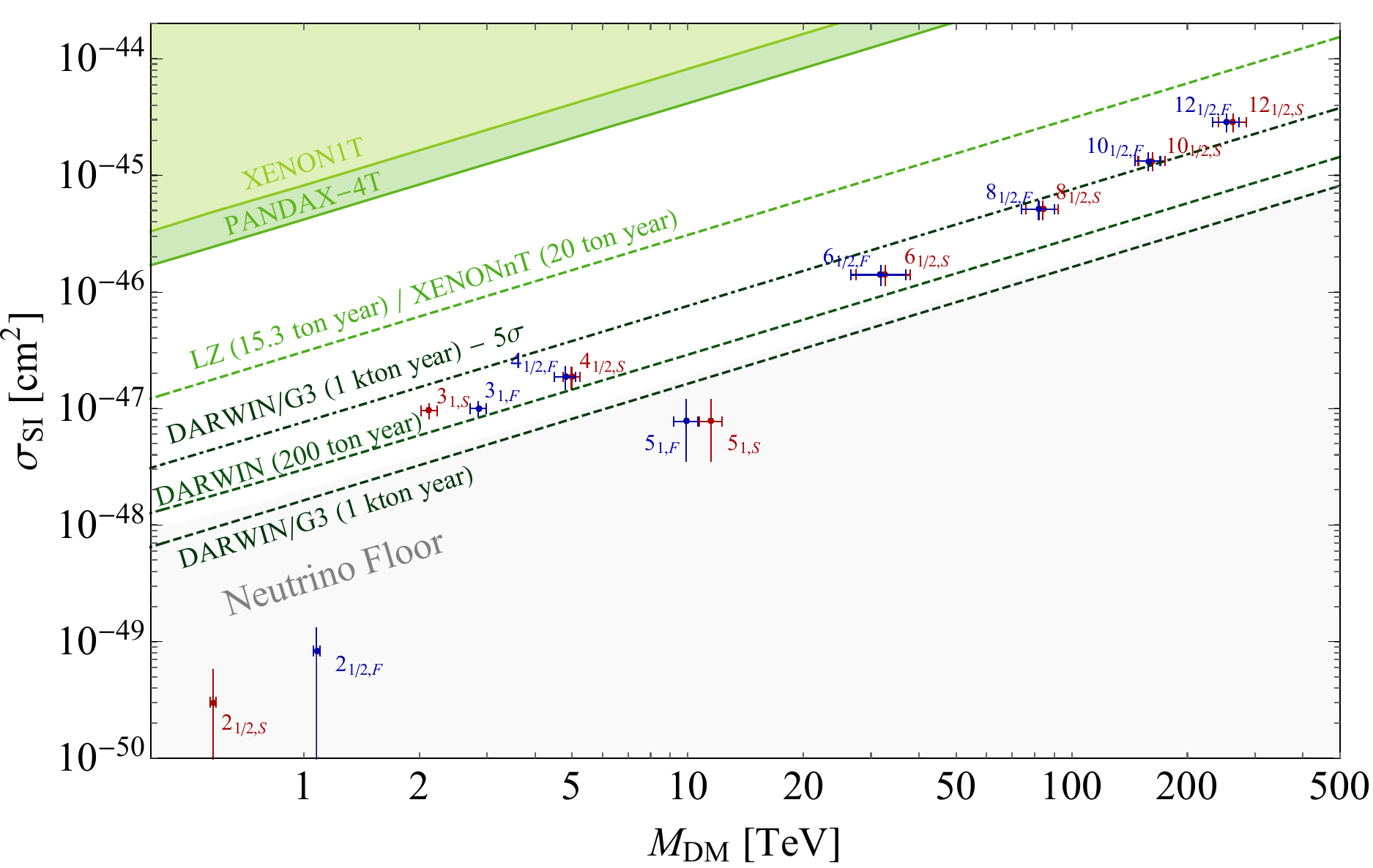}
\caption{In dark green, we display the current constraints from XENON-1T~\cite{XENON:2018voc} and PandaX-4T~\cite{PandaX-4T:2021bab}, while the green dashed line represents the reach of LZ~\cite{Mount:2017qzi}, and the brown green dot-dashed line denotes the ultimate reach of DARWIN~\cite{DARWIN:2016hyl}. The light gray region depicts the neutrino floor for a 200 ton/year exposure, as derived in Ref.~\cite{Billard:2013qya}. {\bf Left:} Here, we present the expected SI cross-section for Majorana $n$-plets (in red) and for real scalar $n$-plets (in blue), assuming no Higgs portal coupling ($\lambda_H=0$). {\bf Right:} This panel showcases the SI cross-section for Dirac fermion $n$-plets (in red) and for complex scalar $n$-plets (in blue) within the minimal splitting scenario.}
\label{fig:dd}
\end{figure}

\medskip
In general, the spin-independent scattering cross-section $\sigma_\text{SI}$ of dark matter on nuclei stems from two main contributions: $i)$ From purely EW loop diagrams, applicable to all EW multiplets; $ii)$ From Higgs-mediated tree-level diagrams, generated by both $\mathcal O_0$ and $\mathcal O_+$ for complex WIMPs. For scenarios with minimal splitting, defined when mass splittings are chosen to be the smallest allowed by the requirements outlined in Sec.~\ref{sec:EWmultiplets}, Higgs-mediated scattering becomes subdominant. In this case, $\sigma_{\text{SI}}$ can be computed solely by considering EW loop diagrams, as for the real multiplets.

Following Refs.~\cite{Hisano:2011cs,Cirelli:2013ufw}, the Lagrangian governing spin-independent DM interactions with quarks and gluons is given by:
\begin{equation}
\label{eq:nogamma}
\mathscr{L}_{\text{eff}}^{\rm{SI}} = f_q^{\rm{EW}} m_q \bar{\chi} \chi \bar{q} q + \frac{g_q^{\rm{EW}}}{\mdm} \bar{\chi} i \partial^{\mu} \gamma^{\nu} \chi \mathcal{O}^q_{\mu\nu} + f_G^{\rm{EW}} \bar{\chi} \chi G_{\mu\nu}G^{\mu\nu},
\end{equation}
where $\mathcal{O}^q_{\mu\nu}$ denotes twist-2 operator of quark. The Wilson coefficients $f_q^{\rm{EW}}$, $f_G^{\rm{EW}}$, and $g_q^{\rm{EW}}$ are expressed as functions of $n$ and $Y$, with detailed expressions provided in~\cite{Hisano:2011cs}. In the non-relativistic limit and dressing Eq.~\eqref{eq:nogamma} at the nucleon scale~\cite{Cirelli:2013ufw}, $\sigma_\text{SI}$ per nucleon (for $\mdm \gg m_N$) can be represented as:
\begin{equation}
\label{elasticSIDD}
\sigma_{\text{SI}} \simeq \frac{4}{\pi} m_N^4 \vert k_N^{\rm{EW}} \vert^2,
\end{equation}
where $m_N$ is the nucleon mass and 
\begin{equation*}
\label{eq:kEW}
k_N^{\rm{EW}} =\!\!\!\!  \sum_{q=u,d,s} f_q^{\rm{EW}} f_{Tq} + \frac{3}{4} (q(2) + \bar{q}(2)) g_q^{\rm{EW}} - \frac{8 \pi}{9 \alpha_s} f_{TG} f_G^{\rm{EW}}\, .
\end{equation*}
Here, the nucleon form factors are defined as $ f_{Tq} = \langle N| m_q \bar{q} q |N\rangle / m_N$, $f_{TG} = 1-\sum_{q=u,d,s} f_{Tq}$, and $\langle N(p)| \mathcal{O}^q_{\mu\nu}|N(p)\rangle = (p_{\mu} p_{\nu}-\frac{1}{4}m_N^2 g_{\mu\nu})(q(2) + \bar{q}(2))/m_N$, where $q(2)$ and $\bar{q}(2)$ are the second moments of the parton distribution functions for a quark or antiquark inside the nucleon~\cite{Hisano:2011cs}. The values of these form factors are obtained from direct computations on the lattice, as reported by the FLAG Collaboration~\cite{FlavourLatticeAveragingGroupFLAG:2021npn} in the case of $N_f=2+1+1$ dynamical quarks~\cite{Alexandrou:2014sha, Freeman:2012ry}.

\smallskip
Fig.~\ref{fig:dd} illustrates the direct detection reach. The left panel displays results for real representations, while the right panel showcases those for complex representations in the minimal splitting scenario. As depicted in the figure, all real representations and most of the complex ones lie above the neutrino floor and are within the detection reach of forthcoming experiments like DARWIN, with the exception of the $2_{1/2}$ and $5_1$ cases, which are discussed below.

A useful  parametric expression for $\sigma_{\text{SI}}$ is
\begin{equation}
\label{eq:cancellation}
    \sigma_{\text{SI}} \approx 10^{-49} \text{ cm}^2 (n^2-1-\xi~ Y^2)^2,
\end{equation}
where $\xi=16.6\pm 1.3$, with the error stemming from the lattice determination of the nucleon form factors. This formula highlights the potential for significant cancellations relative to the natural size of the elastic cross section when $Y\simeq\sqrt{(n^2-1)/\xi}$. Notably, for $n=2$, exact cancellation occurs at $Y\simeq 0.44\pm 0.02$, closely matching the exact hypercharge of the Higgsino. Similarly, cancellation occurs at $Y\simeq 1.2$ for $n=5$. This explain why the signal lies entirely below the neutrino floor in the  $2_{1/2}$ and $5_1$ cases.

\medskip
Regarding spin-dependent (SD) interactions of dark matter with nuclei, these are also induced by EW loops and can yield a larger cross section compared to the spin-independent (SI) one \cite{Hisano:2011cs}. However, the predicted SD cross section for all complex WIMPs consistently falls well below the neutrino floor, making it impossible to test even at future direct detection experiments.

\subsection{Indirect detection}\label{sec:ID}
The current and upcoming ground-based Cherenkov telescopes are exceptionally well-positioned to probe heavy WIMP $n$-plets. This is due to the significant non-perturbative effects, such as Sommerfeld enhancement (SE) and bound state formation (BSF), which greatly enhance the annihilation cross-section across all EW gauge bosons in low-velocity environments. Importantly, these non-perturbative corrections to the annihilation cross-section are more pronounced in low-velocity environments, implying that the annihilation cross-section in the local universe exceeds that in the early universe.

\begin{figure}
\centering
\includegraphics[width=0.95\textwidth]{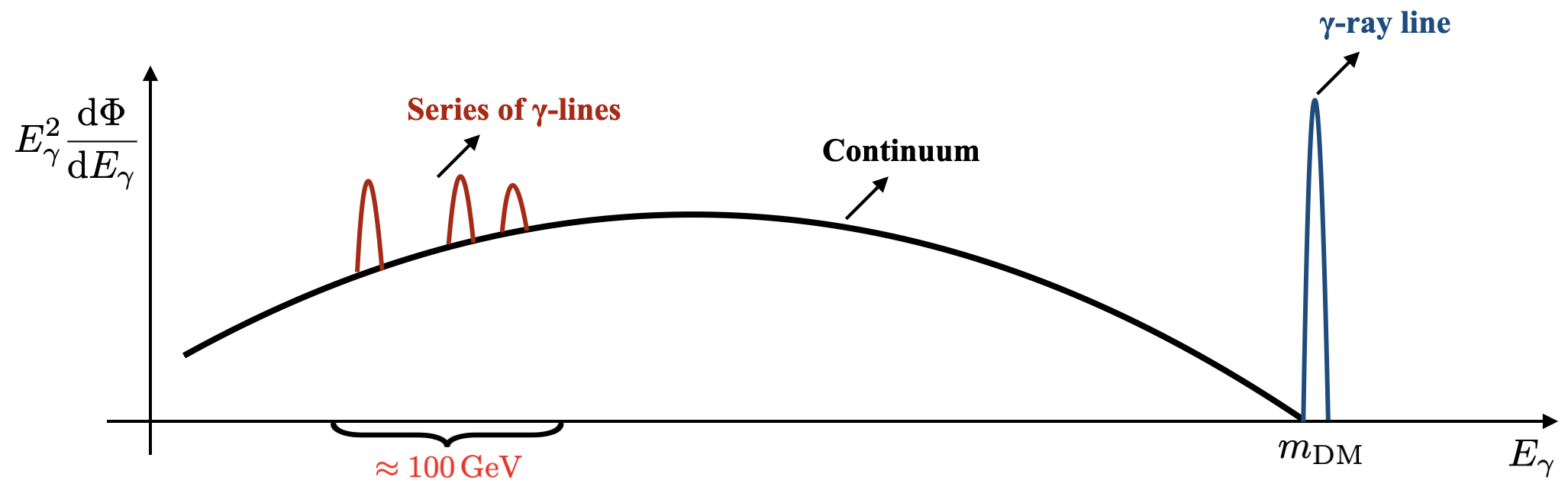}
\caption{An illustrative plot showcases the typical $\gamma$-ray spectrum coming from the annihilation of EW multiplets in astrophysical settings. Beyond the continuum emissions stemming from the decay and hadronization of heavy EW gauge bosons, several distinctive features emerge: notably, a prominent $\gamma$-ray line situated at the spectrum's endpoint, along with a potential array of $\gamma$-ray lines in the energy range of hundreds of GeV, attributable to DM bound-state formation.}
\label{fig:IDflux}
\end{figure}

\smallskip
Fig.~\ref{fig:IDflux} provides an illustrative depiction of the typical $\gamma$-ray flux expected from annihilating EW multiplets in astrophysical targets. This flux exhibits several distinctive features. At the forefront, alongside the continuum arising from the decay and hadronization of heavy EW gauge bosons, is a prominent loop-induced line at the spectrum's endpoint, significantly amplified by the Sommerfeld effect. The cross-section in this channel experiences a substantial boost from the SE, as documented in, for instance, \cite{Cirelli:2007xd, Cirelli:2015bda, Garcia-Cely:2015dda, HESS:2018kom}, and may surpass the gamma-ray continuum originating from the showering, hadronization, and decays of electroweak gauge bosons \cite{Cirelli:2010xx}. Additionally, a series of lines in the energy range of hundreds of GeV is anticipated from the formation of WIMP bound states. In summary, heavy EW multiplets behave akin to atoms emitting gamma-rays, allowing for the exploration of correlations between multiple lines.

From an astrophysical standpoint, the effectiveness of high-energy gamma-ray line searches heavily hinges on the portion of the sky targeted by telescopes. Achieving the optimal choice entails striking a balance between maximizing photon flux at Earth and controlling systematic uncertainties. Two extensively studied astrophysical targets are the Galactic Center (GC)~\cite{Lefranc:2016fgn, Rinchiuso:2018ajn} and the Milky Way's dwarf Spheroidal galaxies (dSphs)~\cite{Lefranc:2016fgn}. In the GC, uncertainties are primarily driven by the importance of baryonic physics in the innermost region of the Milky Way, compounded by the limited knowledge of the dark matter distribution at its center~\cite{Iocco:2015xga, Wegg:2016jxe, Pato:2015dua}. Conversely, dSphs emerge as very clean environments for high-energy $\gamma$-line searches, with residual effects from systematics related to the determination of astrophysical parameters due to limited stellar tracers~\cite{Lefranc:2016dgx,Ullio:2016kvy}.

\begin{figure}
\centering
\includegraphics[width=0.65\textwidth]{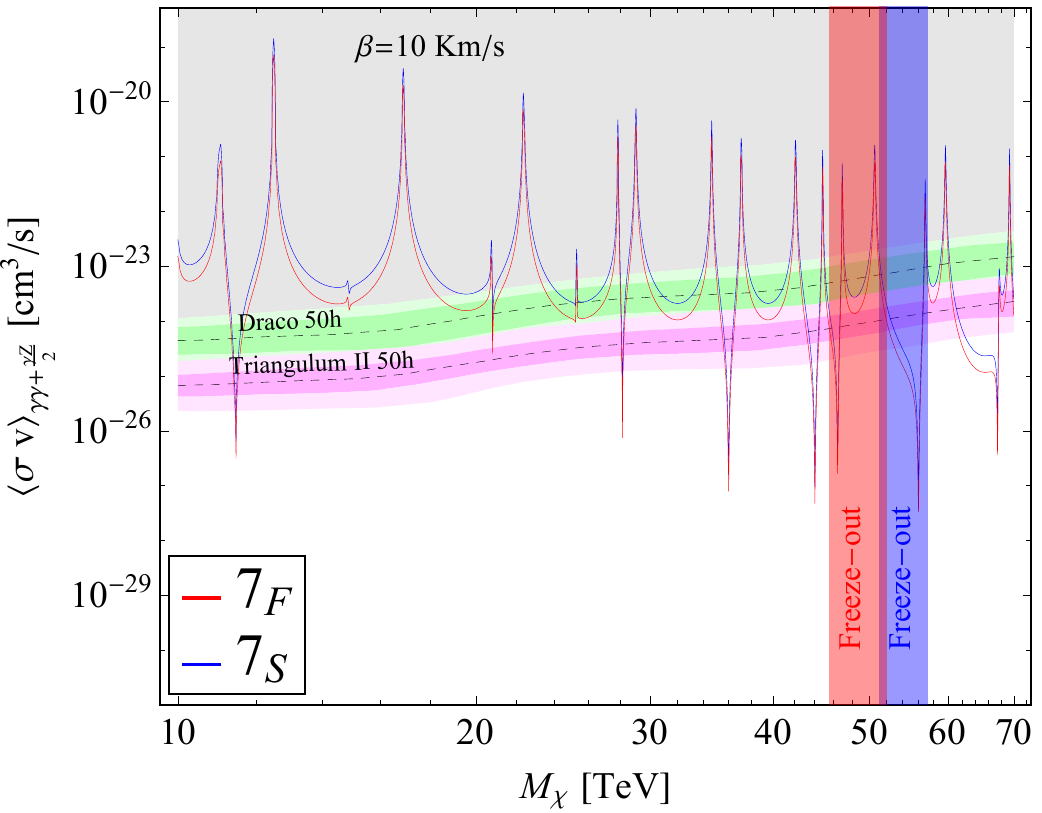}
\caption{We present the expected sensitivities of the Cherenkov Telescope Array (CTA) depicted as dashed black lines, accompanied by the 68\% and 95\% confidence level intervals. These were derived following the methodology outlined in Ref.~\cite{Lefranc:2016fgn}, assuming 50 hours of observation time directed towards Draco (in green) and Triangulum II (in magenta). The SE annihilation cross-sections into the channels contributing to the monochromatic gamma line signal (i.e., $\gamma\gamma$ and $\gamma Z$) are represented for a scalar 7-plet (in blue) and a fermionic 7-plet (in red). Additionally, the vertical bands indicate the predicted thermal masses for the scalar 7-plet (in blue) and the fermionic 7-plet (in red), where the theory uncertainty primarily stems from the neglected next-to-leading order contributions~\cite{Bottaro:2021snn}.}
\label{fig:id}
\end{figure}

\medskip
Motivated by the considerations outlined above, a very preliminary analysis of indirect detection (ID) signals stemming from annihilations of the WIMP 7-plet is presented in Ref.~[citation needed]. The study focuses on the prospects offered by the CTA, considering 50 hours of observation time directed towards two dwarf Spheroidal (dSph) targets in the northern hemisphere: the classic dSph Draco and the ultra-faint one Triangulum II. Fig.~\ref{fig:id} illustrates the Sommerfeld enhancement (SE) annihilation cross-section for the 7-plets at $v=10\text{ km/s}$ overlaid with the experimental reaches of CTA. As depicted, both a 50-hour observation of Triangulum II and of Draco present promising prospects for detecting the high-energy $\gamma$ line in the annihilation spectrum of the 7-plet. 
This analysis is simplified because the signal shape considered essentially consists of a single line at $E_\gamma\simeq M_\chi$. Accordingly, we utilize the CTA prospects derived in Ref.~\cite{Lefranc:2016fgn} for a pure line, disregarding contributions from the continuum spectrum, additional features of the spectral shape induced by the resummation of electroweak (EW) radiation, and the contribution of bound state formation (BSF) to the photon flux. While neglecting BSF may be justified when focusing on very high-energy photons, a meticulous computation of the $\gamma+X$ cross-section, where $X$ denotes any other final state, would be necessary to precisely evaluate the experimental sensitivity~\cite{Baumgart:2017nsr}. 

\medskip
Despite our current inability to make conclusive statements due to significant theoretical uncertainties, it is evident that large $n$-plets represent an ideal target for future Cherenkov telescopes, warranting further theoretical investigation. A complementary open phenomenological question pertains to whether the low-energy gamma lines at $E_\gamma\simeq E_B$ associated with BSF can be effectively distinguished from the continuum (see \cite{Mitridate:2017izz,Mahbubani:2019pij,Baumgart:2023pwn} for preliminary work in this direction). A similar question arises for monochromatic neutrinos from BS annihilations.

\subsection{Colliders}\label{sec:Colliders}
Now, let's explore the potential detection strategies for the direct production of WIMPs at collider experiments. As discussed in Sec.~\ref{sec:WIMPmass}, it becomes apparent that DM masses of $\gtrsim 50$ TeV are necessary to achieve thermal freeze-out for EW multiplets with $n > 5$. Pair-production of these states would demand center-of-mass energies exceeding 100 TeV, a threshold unlikely to be reached at any foreseeable future facility. Conversely, multiplets with $n\leq 5$ exhibit thermal masses in the few TeV range, potentially within the reach of present and future colliders.

The direct reach for these DM candidates at hadron colliders is constrained by the absence of QCD interactions, limiting the production of DM candidates solely to EW interactions. Consequently, the limits at the LHC remain distant from the interesting thermal mass targets. Only a future $pp$ collider operating at collisions around 100 TeV may offer reach for some low-$n$ candidates if such energies can be achieved~\cite{Cirelli:2014dsa,Franceschini:2021aqd}. 

\medskip
On the other hand, lepton colliders primarily achieve reach through indirect effects, such as the modification of angular distributions in simple $f\bar{f}$ production, particularly at center-of-mass energies below the threshold required to produce the DM pair. In this scenario, the reach extends up to masses a few factors above the center-of-mass energy (see e.g.~\cite{DiLuzio:2018jwd}). A high-energy lepton collider, like a muon collider, offers an optimal setting to search for WIMPs. With its substantial center-of-mass energy, clean collision environment, and the ability to produce weakly interacting particles up to kinematical threshold, it stands out as an ideal machine. Specifically, we examine the potential of a future muon collider with a center-of-mass energy exceeding 10 TeV, alongside the baseline integrated luminosity outlined in~\cite{Delahaye:2019omf}
\begin{equation}
\mathcal{L}\simeq 10\;\textrm{ab}^{-1} \cdot \left(  \frac{\sqrt{s}}{10\;\textrm{ TeV}}  \right)^{2} \label{eq:lumi}.
\end{equation}
While the realization of such a machine is presently unattainable, ongoing efforts aim to surmount the associated technological challenges. Early investigations into machine performances~\cite{Boscolo:2018ytm,Palmer:2014nza} have demonstrated the achievability of the luminosity equation \eqref{eq:lumi} for $\sqrt{s}\lesssim 6$ TeV. Further advancements are underway to extend this capability to higher energies.

\medskip
We explore various search channels for low-$n$ EW multiplets and determine the minimum center-of-mass energy and luminosity necessary for directly probing the freeze-out predictions. Initially, we delve into the prospects of observing DM as an undetected carrier of momentum recoiling against one or more SM objects. This involves a systematic examination of all ``mono-$V$'' channels, where DM recoils against an SM gauge boson $V=\gamma,Z,W$. Additionally, we investigate double vector boson production, referred to as ``di-$V$'' channels, wherein the presence of a second SM gauge boson in the final state could enhance sensitivity.

Furthermore, we analyze the reach of disappearing track searches, which are robust predictions of WIMPs in real EW representations, as discussed in Sec.~\ref{sec:Realmultiplets}, by reinterpreting the results of \cite{Capdevilla:2021fmj}. It is essential to contrast the projections for direct production derived here with similar studies conducted within the context of future high-energy proton machines \cite{Cirelli:2014dsa,Low:2014cba} (limited by partial reconstruction of collision kinematics) or electron-positron machines \cite{Fox:2011fx,Bartels:2012ex} (more effective in hunting for lighter DM candidates due to moderate center-of-mass energy).

\begin{figure}
\begin{centering}
\includegraphics[width=0.46\textwidth]{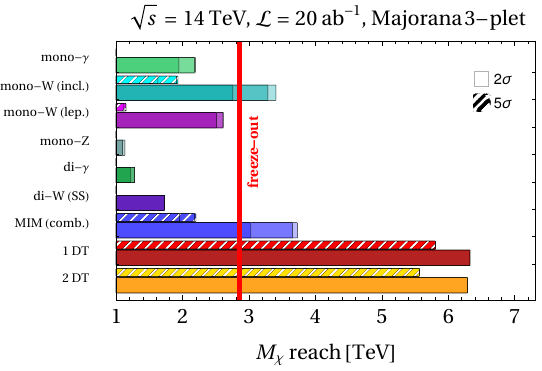}  
\includegraphics[width=0.46\textwidth]{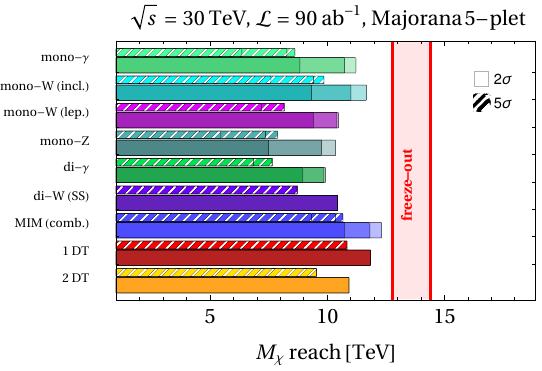}
 \includegraphics[width=0.46\textwidth]{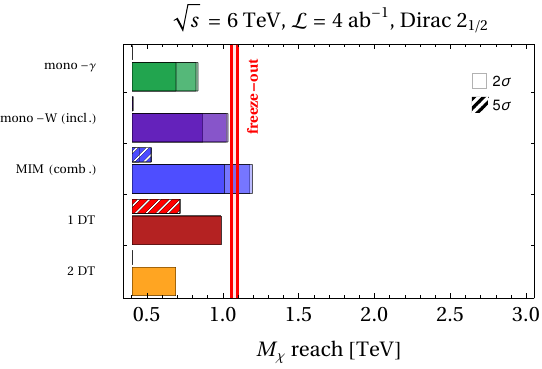}\qquad \, \, 
 \includegraphics[width=0.46\textwidth]{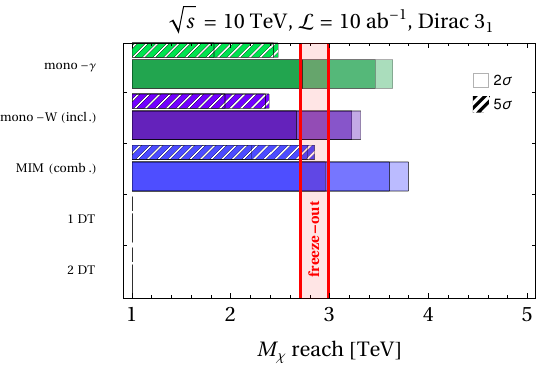}
\end{centering}
\caption{Different bars represent the $2\sigma$ (solid wide) and $5\sigma$ (hatched thin) reach on the WIMP mass at a muon collider across various search channels. The vertical red bands indicate the freeze-out prediction. The first seven(three) bars illustrate the main channels where real(complex) WIMPs would manifest as missing invariant mass (MIM), recoiling against one or more Standard Model (SM) objects: mono-photon, inclusive mono-$W$, leptonic mono-$W$, mono-$Z$, di-photon, same-sign di-$W$, and the combination of all these MIM channels (blue). The last two bars display the reach of disappearing tracks, requiring at least 1 disappearing track (red) or at least 2 tracks (orange). All results assume systematic uncertainties of 0 (light), 0.1\% (medium), or 1\% (dark).  {\bf Top panels:} Majorana 3-plets for $\sqrt{s} = 14 \, {\rm TeV}$ and $\mathcal{L} = 20\,{\rm ab}^{-1}$ on the left, and Majorana 5-plets for $\sqrt{s} = 30\,{\rm TeV}$ and $\mathcal{L} = 90\,{\rm ab}^{-1}$ on the right. {\bf Bottom panels:} Dirac $2_{1/2}$ for $\sqrt{s} = 6\,{\rm TeV}$ and $\mathcal{L} = 4\,{\rm ab}^{-1}$ on the left, and Dirac $3_1$ for $\sqrt{s} = 10\,{\rm TeV}$ and $\mathcal{L} = 10\,{\rm ab}^{-1}$ on the right. \label{fig:barchart}}
\end{figure}

\subsubsection{WIMPs as missing momentum}
We provide a comprehensive analysis of the different channels designed to detect DM as an undetected carrier of momentum. The overarching strategy involves measuring a hard SM particle or a set of particles $X$, which recoil against a pair of invisible objects,
\begin{equation}
    \ell^+\ell^- \to \chi^i \chi^{j} + X \, .\label{eq:monoVgen}
\end{equation}
We consider all components $\chi^i$ of the electroweak (EW) multiplet as invisible, assuming that the soft decay products of the charged states go undetected. Additionally, \eqref{eq:monoVgen} implies the presence of additional soft Standard Model (SM) radiation. The potential of the ``mono-photon'' topology at a future muon collider has been previously explored in~\cite{Han:2020uak}. Here, we aim to enhance this analysis by expanding the range of SM objects recoiling against the invisible DM multiplets.

\medskip
In Fig.~\ref{fig:barchart}, the different bars illustrate the $2\sigma$ (solid wide) and $5\sigma$ (hatched thin) reach on the WIMP mass at a muon collider across various search channels. The vertical red bands represent the freeze-out prediction. The first seven(three) bars depict the primary channels where real(complex) WIMPs would manifest as missing invariant mass (MIM), recoiling against one or more SM objects: mono-photon, inclusive mono-$W$, leptonic mono-$W$, mono-$Z$, di-photon, same-sign di-$W$, and the combination of all these MIM channels (blue). Results are shown assuming systematic uncertainties of 0 (light), 0.1\% (medium), or 1\% (dark).

The top panels present results for real WIMPs: Majorana 3-plets for $\sqrt{s} = 14 \, {\rm TeV}$ and $\mathcal{L} = 20\,{\rm ab}^{-1}$ on the left, and Majorana 5-plets for $\sqrt{s} = 30\,{\rm TeV}$ and $\mathcal{L} = 90\,{\rm ab}^{-1}$ on the right. Notably, the inclusive mono-$W$ channel yields the strongest exclusion for both Majorana 3-plets and 5-plets. The primary impact of the di-boson searches is the reduction of systematic uncertainties in the combined reach of all Missing Mass channels. To probe a Majorana fermion with $n=3$ ($n=5$) using missing-mass searches, a collider with at least $\sqrt{s}\sim 12$~TeV ($\sqrt{s}\sim 35$~TeV) and the baseline integrated luminosity in Eq.~\eqref{eq:lumi} would be required.

In the bottom panels, we present the results for complex WIMPs with $Y \neq 0$: Dirac $2_{1/2}$ for $\sqrt{s} = 6\,{\rm TeV}$ and $\mathcal{L} = 4\,{\rm ab}^{-1}$ on the left, and Dirac $3_1$ for $\sqrt{s} = 10\,{\rm TeV}$ and $\mathcal{L} = 10\,{\rm ab}^{-1}$ on the right. Notably, both the mono-photon and mono-$W$ channels demonstrate similar reach. Specifically, to probe the higgsino (Dirac $2_{1/2}$) using missing-mass searches, a collider with at least $\sqrt{s}\sim 6$ TeV and the baseline integrated luminosity in Eq~\eqref{eq:lumi} is required.

\subsubsection{Disappearing tracks}
Another avenue to probe the production of EW WIMPs at colliders is through the detection of tracks arising from the charged states within the $n$-plet. Specifically, in our benchmark scenarios, the DM is part of a multiplet that also includes charged states. When these charged states are produced in high-energy muon collisions, they decay into DM (via processes like $\chi^{\pm}\to \chi^0 \pi^{\pm}$) inside the detector. If the charged states are sufficiently long-lived, one can search for disappearing tracks (DTs) of these charged particles to isolate the DM signal from the background, primarily composed of neutrinos. These resulting tracks are often too short for conventional track reconstruction methods to be effective. They manifest as disappearing tracks, characterized by missing hits in the outermost layers of the tracker and minimal to no activity in the calorimeter and the muon chamber. Higher electric charge states in larger multiplets decay promptly to $\chi^{\pm}$, thereby contributing to the number of disappearing tracks.

\medskip
We investigate mono-photon events accompanied by disappearing tracks and explore event-selection strategies aimed at identifying a WIMP signal, following the methodology outlined in~\cite{Capdevilla:2021fmj}. Two distinct event-selection strategies are considered: $i)$ Events featuring at least one disappearing track with transverse momentum $p_{\mathrm{T}}>300 \text{ GeV}$, accompanied by a hard photon with energy $E_\gamma>25 \text{ GeV}$; $ii)$ Events containing a hard photon along with two disappearing tracks originating from the same point along the beam axis. To assess the potential reach, we adopt a cut-and-count approach and neglect systematic uncertainties. Additional details are provided in the Appendix of~\cite{Bottaro:2021snn, Bottaro:2022one} for completeness.

\smallskip
It is crucial to emphasize that for real representations with $Y=0$, the mass splitting is solely determined by gauge interactions, leading to a precise prediction for the lifetime of the singly-charged component. These lifetimes have been explicitly computed in Refs.~\cite{Ibe:2012sx,McKay:2017xlc}:
\begin{equation}
c\tau_{\chi^+}\simeq \frac{480\text{ mm}}{(n^2-1)}\ .
\end{equation}
For instance, in the case of the Majorana 3-plet, we anticipate tracks of approximately half a centimeter, which are in principle detectable. For complex representations with $Y \neq 0$, the mass splitting is determined solely by gauge interactions only for multiplets with maximal hypercharge and is typically larger than that obtained for real representations. In Fig.~4 of~\cite{Bottaro:2022one}, the significance of detecting 1 DT and 2 DT as a function of the track length is illustrated for various complex WIMPs. Generally, if the mass splitting is solely determined by gauge interactions, the track length is too short to be detectable. The only plausible approach to observe a signal is to manually reduce the EW mass splitting using the higher-dimensional operator $\ochgd$; however, this option involves fine-tuning.
	
\medskip
The results of our recast are depicted in the last two columns of Fig.~\ref{fig:barchart}. It can be observed that DTs exhibit significant efficacy, particularly in the case of the 3-plet, where the reach extends nearly up to the kinematic threshold. Specifically, an EW 3-plet WIMP with a mass predicted by thermal freeze-out could be easily discovered at a $6$ TeV muon collider, as also suggested in \cite{Capdevilla:2021fmj,Han:2020uak}. However, for higher real $n$-plets and typically for complex WIMPs with purely EW mass splittings, the DT method substantially loses its exclusion power due to shorter lifetimes of the $\chi^{\pm}\to \chi^0 \pi^{\pm}$ decays. In the case of the real 5-plet, the DT reach is comparable to the combined reach of the MIM searches. Conversely, for the complex representations illustrated in the plot, DTs are entirely subleading.

\section{Conclusions and outlooks}
In conclusion, this brief review has outlined the phenomenology of EW multiplets as candidates for DM, serving as a prototype for WIMP DM. The necessity of thermal production implies multi-TeV mass scales, and the extensive phenomenology of these candidates remains largely unexplored.

\medskip
Looking ahead, we anticipate a trajectory over the next 30 years that leads to a deeper understanding of DM with purely EW interactions. This journey will benefit from the combined efforts of cosmological probes and the potential realization of a muon collider. Ground-based Cherenkov telescopes (like CTA), in particular, are poised to make significant contributions by probing heavy WIMP $n$-plets, due to the enhanced annihilation cross-section in low-velocity environments. The upcoming generation of kton-scale direct detection experiments, such as DARWIN, holds great promise for detecting most real representations and many complex ones. However, challenges remain for certain cases, such as the $2_{1/2}$ and $5_1$ multiplets. To probe low-$n$ multiplets like the supersymmetric higgsino (Dirac $2_{1/2}$) and the Wino (Majorana $3_0$), a 14 TeV muon collider with benchmark luminosity would be necessary.


\begin{thebibliography}{99}

%\cite{Planck:2018vyg}
\bibitem{Planck:2018vyg}
N.~Aghanim \textit{et al.} [Planck],
%``Planck 2018 results. VI. Cosmological parameters,''
Astron. Astrophys. \textbf{641} (2020), A6
[erratum: Astron. Astrophys. \textbf{652} (2021), C4]
doi:10.1051/0004-6361/201833910
[arXiv:1807.06209 [astro-ph.CO]].
%13632 citations counted in INSPIRE as of 08 May 2024

%\cite{Bottaro:2021snn}
\bibitem{Bottaro:2021snn}
S.~Bottaro, D.~Buttazzo, M.~Costa, R.~Franceschini, P.~Panci, D.~Redigolo and L.~Vittorio,
%``Closing the window on WIMP Dark Matter,''
Eur. Phys. J. C \textbf{82} (2022) no.1, 31
doi:10.1140/epjc/s10052-021-09917-9
[arXiv:2107.09688 [hep-ph]].
%80 citations counted in INSPIRE as of 08 May 2024

%\cite{Bottaro:2022one}
\bibitem{Bottaro:2022one}
S.~Bottaro, D.~Buttazzo, M.~Costa, R.~Franceschini, P.~Panci, D.~Redigolo and L.~Vittorio,
%``The last complex WIMPs standing,''
Eur. Phys. J. C \textbf{82} (2022) no.11, 992
doi:10.1140/epjc/s10052-022-10918-5
[arXiv:2205.04486 [hep-ph]].
%38 citations counted in INSPIRE as of 08 May 2024

%\cite{Wells:2003tf}
\bibitem{Wells:2003tf}
J.~D.~Wells,
%``Implications of supersymmetry breaking with a little hierarchy between gauginos and scalars,''
[arXiv:hep-ph/0306127 [hep-ph]].
%222 citations counted in INSPIRE as of 08 May 2024

%\cite{Arkani-Hamed:2004ymt}
\bibitem{Arkani-Hamed:2004ymt}
N.~Arkani-Hamed and S.~Dimopoulos,
%``Supersymmetric unification without low energy supersymmetry and signatures for fine-tuning at the LHC,''
JHEP \textbf{06} (2005), 073
doi:10.1088/1126-6708/2005/06/073
[arXiv:hep-th/0405159 [hep-th]].
%1301 citations counted in INSPIRE as of 08 May 2024

%\cite{Giudice:2004tc}
\bibitem{Giudice:2004tc}
G.~F.~Giudice and A.~Romanino,
%``Split supersymmetry,''
Nucl. Phys. B \textbf{699} (2004), 65-89
[erratum: Nucl. Phys. B \textbf{706} (2005), 487-487]
doi:10.1016/j.nuclphysb.2004.08.001
[arXiv:hep-ph/0406088 [hep-ph]].
%1075 citations counted in INSPIRE as of 08 May 2024

%\cite{DEramo:2014urw}
\bibitem{DEramo:2014urw}
F.~D'Eramo, L.~J.~Hall and D.~Pappadopulo,
%``Multiverse Dark Matter: SUSY or Axions,''
JHEP \textbf{11} (2014), 108
doi:10.1007/JHEP11(2014)108
[arXiv:1409.5123 [hep-ph]].
%22 citations counted in INSPIRE as of 08 May 2024

%\cite{Cirelli:2005uq}
\bibitem{Cirelli:2005uq}
M.~Cirelli, N.~Fornengo and A.~Strumia,
%``Minimal dark matter,''
Nucl. Phys. B \textbf{753} (2006), 178-194
doi:10.1016/j.nuclphysb.2006.07.012
[arXiv:hep-ph/0512090 [hep-ph]].
%998 citations counted in INSPIRE as of 08 May 2024

%\cite{Cirelli:2007xd}
\bibitem{Cirelli:2007xd}
M.~Cirelli, A.~Strumia and M.~Tamburini,
%``Cosmology and Astrophysics of Minimal Dark Matter,''
Nucl. Phys. B \textbf{787} (2007), 152-175
doi:10.1016/j.nuclphysb.2007.07.023
[arXiv:0706.4071 [hep-ph]].
%609 citations counted in INSPIRE as of 08 May 2024

%\cite{Cirelli:2009uv}
\bibitem{Cirelli:2009uv}
M.~Cirelli and A.~Strumia,
%``Minimal Dark Matter: Model and results,''
New J. Phys. \textbf{11} (2009), 105005
doi:10.1088/1367-2630/11/10/105005
[arXiv:0903.3381 [hep-ph]].
%248 citations counted in INSPIRE as of 08 May 2024

%\cite{Hambye:2009pw}
\bibitem{Hambye:2009pw}
T.~Hambye, F.~S.~Ling, L.~Lopez Honorez and J.~Rocher,
%``Scalar Multiplet Dark Matter,''
JHEP \textbf{07} (2009), 090
[erratum: JHEP \textbf{05} (2010), 066]
doi:10.1007/JHEP05(2010)066
[arXiv:0903.4010 [hep-ph]].
%308 citations counted in INSPIRE as of 08 May 2024

%\cite{DelNobile:2015bqo}
\bibitem{DelNobile:2015bqo}
E.~Del Nobile, M.~Nardecchia and P.~Panci,
%``Millicharge or Decay: A Critical Take on Minimal Dark Matter,''
JCAP \textbf{04} (2016), 048
doi:10.1088/1475-7516/2016/04/048
[arXiv:1512.05353 [hep-ph]].
%72 citations counted in INSPIRE as of 08 May 2024

%\cite{Cheng:1998hc}
\bibitem{Cheng:1998hc}
H.~C.~Cheng, B.~A.~Dobrescu and K.~T.~Matchev,
%``Generic and chiral extensions of the supersymmetric standard model,''
Nucl. Phys. B \textbf{543} (1999), 47-72
doi:10.1016/S0550-3213(99)00012-7
[arXiv:hep-ph/9811316 [hep-ph]].
%152 citations counted in INSPIRE as of 08 May 2024

%\cite{Feng:1999fu}
\bibitem{Feng:1999fu}
J.~L.~Feng, T.~Moroi, L.~Randall, M.~Strassler and S.~f.~Su,
%``Discovering supersymmetry at the Tevatron in wino LSP scenarios,''
Phys. Rev. Lett. \textbf{83} (1999), 1731-1734
doi:10.1103/PhysRevLett.83.1731
[arXiv:hep-ph/9904250 [hep-ph]].
%267 citations counted in INSPIRE as of 08 May 2024

%\cite{Gherghetta:1999sw}
\bibitem{Gherghetta:1999sw}
T.~Gherghetta, G.~F.~Giudice and J.~D.~Wells,
%``Phenomenological consequences of supersymmetry with anomaly induced masses,''
Nucl. Phys. B \textbf{559} (1999), 27-47
doi:10.1016/S0550-3213(99)00429-0
[arXiv:hep-ph/9904378 [hep-ph]].
%477 citations counted in INSPIRE as of 08 May 2024

%\cite{Goodman:1984dc}
\bibitem{Goodman:1984dc}
M.~W.~Goodman and E.~Witten,
%``Detectability of Certain Dark Matter Candidates,''
Phys. Rev. D \textbf{31} (1985), 3059
doi:10.1103/PhysRevD.31.3059
%1511 citations counted in INSPIRE as of 08 May 2024

%\cite{Cassel:2009wt}
\bibitem{Cassel:2009wt}
S.~Cassel,
%``Sommerfeld factor for arbitrary partial wave processes,''
J. Phys. G \textbf{37} (2010), 105009
doi:10.1088/0954-3899/37/10/105009
[arXiv:0903.5307 [hep-ph]].
%268 citations counted in INSPIRE as of 08 May 2024

%\cite{Mitridate:2017izz}
\bibitem{Mitridate:2017izz}
A.~Mitridate, M.~Redi, J.~Smirnov and A.~Strumia,
%``Cosmological Implications of Dark Matter Bound States,''
JCAP \textbf{05} (2017), 006
doi:10.1088/1475-7516/2017/05/006
[arXiv:1702.01141 [hep-ph]].
%131 citations counted in INSPIRE as of 08 May 2024

%\cite{Harz:2018csl}
\bibitem{Harz:2018csl}
J.~Harz and K.~Petraki,
%``Radiative bound-state formation in unbroken perturbative non-Abelian theories and implications for dark matter,''
JHEP \textbf{07} (2018), 096
doi:10.1007/JHEP07(2018)096
[arXiv:1805.01200 [hep-ph]].
%68 citations counted in INSPIRE as of 08 May 2024

%\cite{Hisano:2011cs}
\bibitem{Hisano:2011cs}
J.~Hisano, K.~Ishiwata, N.~Nagata and T.~Takesako,
%``Direct Detection of Electroweak-Interacting Dark Matter,''
JHEP \textbf{07} (2011), 005
doi:10.1007/JHEP07(2011)005
[arXiv:1104.0228 [hep-ph]].
%155 citations counted in INSPIRE as of 08 May 2024

%\cite{Cirelli:2013ufw}
\bibitem{Cirelli:2013ufw}
M.~Cirelli, E.~Del Nobile and P.~Panci,
%``Tools for model-independent bounds in direct dark matter searches,''
JCAP \textbf{10} (2013), 019
doi:10.1088/1475-7516/2013/10/019
[arXiv:1307.5955 [hep-ph]].
%298 citations counted in INSPIRE as of 08 May 2024

%\cite{XENON:2018voc}
\bibitem{XENON:2018voc}
E.~Aprile \textit{et al.} [XENON],
%``Dark Matter Search Results from a One Ton-Year Exposure of XENON1T,''
Phys. Rev. Lett. \textbf{121} (2018) no.11, 111302
doi:10.1103/PhysRevLett.121.111302
[arXiv:1805.12562 [astro-ph.CO]].
%2201 citations counted in INSPIRE as of 08 May 2024

%\cite{PandaX-4T:2021bab}
\bibitem{PandaX-4T:2021bab}
Y.~Meng \textit{et al.} [PandaX-4T],
%``Dark Matter Search Results from the PandaX-4T Commissioning Run,''
Phys. Rev. Lett. \textbf{127} (2021) no.26, 261802
doi:10.1103/PhysRevLett.127.261802
[arXiv:2107.13438 [hep-ex]].
%398 citations counted in INSPIRE as of 08 May 2024

%\cite{Mount:2017qzi}
\bibitem{Mount:2017qzi}
B.~J.~Mount, S.~Hans, R.~Rosero, M.~Yeh, C.~Chan, R.~J.~Gaitskell, D.~Q.~Huang, J.~Makkinje, D.~C.~Malling and M.~Pangilinan, \textit{et al.}
%``LUX-ZEPLIN (LZ) Technical Design Report,''
[arXiv:1703.09144 [physics.ins-det]].
%254 citations counted in INSPIRE as of 08 May 2024

%\cite{DARWIN:2016hyl}
\bibitem{DARWIN:2016hyl}
J.~Aalbers \textit{et al.} [DARWIN],
%``DARWIN: towards the ultimate dark matter detector,''
JCAP \textbf{11} (2016), 017
doi:10.1088/1475-7516/2016/11/017
[arXiv:1606.07001 [astro-ph.IM]].
%709 citations counted in INSPIRE as of 08 May 2024

%\cite{Billard:2013qya}
\bibitem{Billard:2013qya}
J.~Billard, L.~Strigari and E.~Figueroa-Feliciano,
%``Implication of neutrino backgrounds on the reach of next generation dark matter direct detection experiments,''
Phys. Rev. D \textbf{89} (2014) no.2, 023524
doi:10.1103/PhysRevD.89.023524
[arXiv:1307.5458 [hep-ph]].
%765 citations counted in INSPIRE as of 08 May 2024

%\cite{FlavourLatticeAveragingGroupFLAG:2021npn}
\bibitem{FlavourLatticeAveragingGroupFLAG:2021npn}
Y.~Aoki \textit{et al.} [Flavour Lattice Averaging Group (FLAG)],
%``FLAG Review 2021,''
Eur. Phys. J. C \textbf{82} (2022) no.10, 869
doi:10.1140/epjc/s10052-022-10536-1
[arXiv:2111.09849 [hep-lat]].
%589 citations counted in INSPIRE as of 08 May 2024

%\cite{Alexandrou:2014sha}
\bibitem{Alexandrou:2014sha}
C.~Alexandrou, V.~Drach, K.~Jansen, C.~Kallidonis and G.~Koutsou,
%``Baryon spectrum with $N_f=2+1+1$ twisted mass fermions,''
Phys. Rev. D \textbf{90} (2014) no.7, 074501
doi:10.1103/PhysRevD.90.074501
[arXiv:1406.4310 [hep-lat]].
%179 citations counted in INSPIRE as of 08 May 2024

%\cite{Freeman:2012ry}
\bibitem{Freeman:2012ry}
W.~Freeman \textit{et al.} [MILC],
%``Intrinsic strangeness and charm of the nucleon using improved staggered fermions,''
Phys. Rev. D \textbf{88} (2013), 054503
doi:10.1103/PhysRevD.88.054503
[arXiv:1204.3866 [hep-lat]].
%102 citations counted in INSPIRE as of 08 May 2024

%\cite{Cirelli:2015bda}
\bibitem{Cirelli:2015bda}
M.~Cirelli, T.~Hambye, P.~Panci, F.~Sala and M.~Taoso,
%``Gamma ray tests of Minimal Dark Matter,''
JCAP \textbf{10} (2015), 026
doi:10.1088/1475-7516/2015/10/026
[arXiv:1507.05519 [hep-ph]].
%109 citations counted in INSPIRE as of 08 May 2024

%\cite{Garcia-Cely:2015dda}
\bibitem{Garcia-Cely:2015dda}
C.~Garcia-Cely, A.~Ibarra, A.~S.~Lamperstorfer and M.~H.~G.~Tytgat,
%``Gamma-rays from Heavy Minimal Dark Matter,''
JCAP \textbf{10} (2015), 058
doi:10.1088/1475-7516/2015/10/058
[arXiv:1507.05536 [hep-ph]].
%73 citations counted in INSPIRE as of 08 May 2024

%\cite{HESS:2018kom}
\bibitem{HESS:2018kom}
H.~Abdalla \textit{et al.} [HESS],
%``Searches for gamma-ray lines and 'pure WIMP' spectra from Dark Matter annihilations in dwarf galaxies with H.E.S.S,''
JCAP \textbf{11} (2018), 037
doi:10.1088/1475-7516/2018/11/037
[arXiv:1810.00995 [astro-ph.HE]].
%62 citations counted in INSPIRE as of 08 May 2024

%\cite{Cirelli:2010xx}
\bibitem{Cirelli:2010xx}
M.~Cirelli, G.~Corcella, A.~Hektor, G.~Hutsi, M.~Kadastik, P.~Panci, M.~Raidal, F.~Sala and A.~Strumia,
%``PPPC 4 DM ID: A Poor Particle Physicist Cookbook for Dark Matter Indirect Detection,''
JCAP \textbf{03} (2011), 051
[erratum: JCAP \textbf{10} (2012), E01]
doi:10.1088/1475-7516/2012/10/E01
[arXiv:1012.4515 [hep-ph]].
%1081 citations counted in INSPIRE as of 08 May 2024

%\cite{Lefranc:2016fgn}
\bibitem{Lefranc:2016fgn}
V.~Lefranc, E.~Moulin, P.~Panci, F.~Sala and J.~Silk,
%``Dark Matter in $\gamma$ lines: Galactic Center vs dwarf galaxies,''
JCAP \textbf{09} (2016), 043
doi:10.1088/1475-7516/2016/09/043
[arXiv:1608.00786 [astro-ph.HE]].
%50 citations counted in INSPIRE as of 08 May 2024

%\cite{Rinchiuso:2018ajn}
\bibitem{Rinchiuso:2018ajn}
L.~Rinchiuso, N.~L.~Rodd, I.~Moult, E.~Moulin, M.~Baumgart, T.~Cohen, T.~R.~Slatyer, I.~W.~Stewart and V.~Vaidya,
%``Hunting for Heavy Winos in the Galactic Center,''
Phys. Rev. D \textbf{98} (2018) no.12, 123014
doi:10.1103/PhysRevD.98.123014
[arXiv:1808.04388 [astro-ph.HE]].
%30 citations counted in INSPIRE as of 08 May 2024

%\cite{Iocco:2015xga}
\bibitem{Iocco:2015xga}
F.~Iocco, M.~Pato and G.~Bertone,
%``Evidence for dark matter in the inner Milky Way,''
Nature Phys. \textbf{11} (2015), 245-248
doi:10.1038/nphys3237
[arXiv:1502.03821 [astro-ph.GA]].
%173 citations counted in INSPIRE as of 08 May 2024

%\cite{Wegg:2016jxe}
\bibitem{Wegg:2016jxe}
C.~Wegg, O.~Gerhard and M.~Portail,
%``MOA-II Galactic microlensing constraints: the inner Milky Way has a low dark matter fraction and a near maximal disc,''
Mon. Not. Roy. Astron. Soc. \textbf{463} (2016) no.1, 557-570
doi:10.1093/mnras/stw1954
[arXiv:1607.06462 [astro-ph.GA]].
%31 citations counted in INSPIRE as of 08 May 2024

%\cite{Pato:2015dua}
\bibitem{Pato:2015dua}
M.~Pato, F.~Iocco and G.~Bertone,
%``Dynamical constraints on the dark matter distribution in the Milky Way,''
JCAP \textbf{12} (2015), 001
doi:10.1088/1475-7516/2015/12/001
[arXiv:1504.06324 [astro-ph.GA]].
%171 citations counted in INSPIRE as of 08 May 2024

%\cite{Lefranc:2016dgx}
\bibitem{Lefranc:2016dgx}
V.~Lefranc, G.~A.~Mamon and P.~Panci,
%``Prospects for annihilating Dark Matter towards Milky Way's dwarf galaxies by the Cherenkov Telescope Array,''
JCAP \textbf{09} (2016), 021
doi:10.1088/1475-7516/2016/09/021
[arXiv:1605.02793 [astro-ph.HE]].
%29 citations counted in INSPIRE as of 08 May 2024

%\cite{Ullio:2016kvy}
\bibitem{Ullio:2016kvy}
P.~Ullio and M.~Valli,
%``A critical reassessment of particle Dark Matter limits from dwarf satellites,''
JCAP \textbf{07} (2016), 025
doi:10.1088/1475-7516/2016/07/025
[arXiv:1603.07721 [astro-ph.GA]].
%59 citations counted in INSPIRE as of 08 May 2024

%\cite{Baumgart:2017nsr}
\bibitem{Baumgart:2017nsr}
M.~Baumgart, T.~Cohen, I.~Moult, N.~L.~Rodd, T.~R.~Slatyer, M.~P.~Solon, I.~W.~Stewart and V.~Vaidya,
%``Resummed Photon Spectra for WIMP Annihilation,''
JHEP \textbf{03} (2018), 117
doi:10.1007/JHEP03(2018)117
[arXiv:1712.07656 [hep-ph]].
%55 citations counted in INSPIRE as of 08 May 2024

%\cite{Mahbubani:2019pij}
\bibitem{Mahbubani:2019pij}
R.~Mahbubani, M.~Redi and A.~Tesi,
%``Indirect detection of composite asymmetric dark matter,''
Phys. Rev. D \textbf{101} (2020) no.10, 103037
doi:10.1103/PhysRevD.101.103037
[arXiv:1908.00538 [hep-ph]].
%19 citations counted in INSPIRE as of 08 May 2024

%\cite{Baumgart:2023pwn}
\bibitem{Baumgart:2023pwn}
M.~Baumgart, N.~L.~Rodd, T.~R.~Slatyer and V.~Vaidya,
%``The quintuplet annihilation spectrum,''
JHEP \textbf{01} (2024), 158
doi:10.1007/JHEP01(2024)158
[arXiv:2309.11562 [hep-ph]].
%4 citations counted in INSPIRE as of 08 May 2024

%\cite{Cirelli:2014dsa}
\bibitem{Cirelli:2014dsa}
M.~Cirelli, F.~Sala and M.~Taoso,
%``Wino-like Minimal Dark Matter and future colliders,''
JHEP \textbf{10} (2014), 033
[erratum: JHEP \textbf{01} (2015), 041]
doi:10.1007/JHEP01(2015)041
[arXiv:1407.7058 [hep-ph]].
%138 citations counted in INSPIRE as of 08 May 2024

%\cite{Franceschini:2021aqd}
\bibitem{Franceschini:2021aqd}
R.~Franceschini and M.~Greco,
%``Higgs and BSM Physics at the Future Muon Collider,''
Symmetry \textbf{13} (2021) no.5, 851
doi:10.3390/sym13050851
[arXiv:2104.05770 [hep-ph]].
%32 citations counted in INSPIRE as of 08 May 2024

%\cite{DiLuzio:2018jwd}
\bibitem{DiLuzio:2018jwd}
L.~Di Luzio, R.~Gr\"ober and G.~Panico,
%``Probing new electroweak states via precision measurements at the LHC and future colliders,''
JHEP \textbf{01} (2019), 011
doi:10.1007/JHEP01(2019)011
[arXiv:1810.10993 [hep-ph]].
%73 citations counted in INSPIRE as of 08 May 2024

%\cite{Delahaye:2019omf}
\bibitem{Delahaye:2019omf}
J.~P.~Delahaye, M.~Diemoz, K.~Long, B.~Mansouli\'e, N.~Pastrone, L.~Rivkin, D.~Schulte, A.~Skrinsky and A.~Wulzer,
%``Muon Colliders,''
[arXiv:1901.06150 [physics.acc-ph]].
%238 citations counted in INSPIRE as of 08 May 2024

%\cite{Boscolo:2018ytm}
\bibitem{Boscolo:2018ytm}
M.~Boscolo, J.~P.~Delahaye and M.~Palmer,
%``The future prospects of muon colliders and neutrino factories,''
Rev. Accel. Sci. Tech. \textbf{10} (2019) no.01, 189-214
doi:10.1142/9789811209604\_0010
[arXiv:1808.01858 [physics.acc-ph]].
%62 citations counted in INSPIRE as of 08 May 2024

%\cite{Palmer:2014nza}
\bibitem{Palmer:2014nza}
R.~B.~Palmer,
%``Muon Colliders,''
Rev. Accel. Sci. Tech. \textbf{7} (2014), 137-159
doi:10.1142/S1793626814300072
%75 citations counted in INSPIRE as of 08 May 2024

%\cite{Capdevilla:2021fmj}
\bibitem{Capdevilla:2021fmj}
R.~Capdevilla, F.~Meloni, R.~Simoniello and J.~Zurita,
%``Hunting wino and higgsino dark matter at the muon collider with disappearing tracks,''
JHEP \textbf{06} (2021), 133
doi:10.1007/JHEP06(2021)133
[arXiv:2102.11292 [hep-ph]].
%64 citations counted in INSPIRE as of 08 May 2024

%\cite{Low:2014cba}
\bibitem{Low:2014cba}
M.~Low and L.~T.~Wang,
%``Neutralino dark matter at 14 TeV and 100 TeV,''
JHEP \textbf{08} (2014), 161
doi:10.1007/JHEP08(2014)161
[arXiv:1404.0682 [hep-ph]].
%227 citations counted in INSPIRE as of 08 May 2024

%\cite{Fox:2011fx}
\bibitem{Fox:2011fx}
P.~J.~Fox, R.~Harnik, J.~Kopp and Y.~Tsai,
%``LEP Shines Light on Dark Matter,''
Phys. Rev. D \textbf{84} (2011), 014028
doi:10.1103/PhysRevD.84.014028
[arXiv:1103.0240 [hep-ph]].
%371 citations counted in INSPIRE as of 08 May 2024

%\cite{Bartels:2012ex}
\bibitem{Bartels:2012ex}
C.~Bartels, M.~Berggren and J.~List,
%``Characterising WIMPs at a future $e^+e^-$ Linear Collider,''
Eur. Phys. J. C \textbf{72} (2012), 2213
doi:10.1140/epjc/s10052-012-2213-9
[arXiv:1206.6639 [hep-ex]].
%63 citations counted in INSPIRE as of 08 May 2024

%\cite{Han:2020uak}
\bibitem{Han:2020uak}
T.~Han, Z.~Liu, L.~T.~Wang and X.~Wang,
%``WIMPs at High Energy Muon Colliders,''
Phys. Rev. D \textbf{103} (2021) no.7, 075004
doi:10.1103/PhysRevD.103.075004
[arXiv:2009.11287 [hep-ph]].
%98 citations counted in INSPIRE as of 08 May 2024

%\cite{Capdevilla:2021fmj}
\bibitem{Capdevilla:2021fmj}
R.~Capdevilla, F.~Meloni, R.~Simoniello and J.~Zurita,
%``Hunting wino and higgsino dark matter at the muon collider with disappearing tracks,''
JHEP \textbf{06} (2021), 133
doi:10.1007/JHEP06(2021)133
[arXiv:2102.11292 [hep-ph]].
%64 citations counted in INSPIRE as of 08 May 2024

%\cite{Ibe:2012sx}
\bibitem{Ibe:2012sx}
M.~Ibe, S.~Matsumoto and R.~Sato,
%``Mass Splitting between Charged and Neutral Winos at Two-Loop Level,''
Phys. Lett. B \textbf{721} (2013), 252-260
doi:10.1016/j.physletb.2013.03.015
[arXiv:1212.5989 [hep-ph]].
%192 citations counted in INSPIRE as of 08 May 2024

%\cite{McKay:2017xlc}
\bibitem{McKay:2017xlc}
J.~McKay and P.~Scott,
%``Two-loop mass splittings in electroweak multiplets: winos and minimal dark matter,''
Phys. Rev. D \textbf{97} (2018) no.5, 055049
doi:10.1103/PhysRevD.97.055049
[arXiv:1712.00968 [hep-ph]].
%22 citations counted in INSPIRE as of 08 May 2024


\end{thebibliography}
\end{document}